\documentclass[aps,prb,twocolumn,groupedaddress,showpacs,intlimits,amsmath,amssymb,floatfix,citeautoscript,footinbib]{revtex4-1}
\usepackage{bm}
\usepackage[final]{graphicx}
\usepackage[hang,center]{subfigure}

\emergencystretch=\maxdimen
\begin{document}

\title{Renormalization of spectra by phase competition in the half-filled Hubbard-Holstein model }

\author{E. A. Nowadnick$^{1,2,3}$}
\author{S. Johnston$^{4,5}$}
\author{B. Moritz$^{3,6}$}
\author{T. P. Devereaux$^{3}$}
\affiliation{$^1$School of Applied and Engineering Physics, Cornell University, Ithaca, NY 14853 USA}
\affiliation{$^2$Department of Physics, Stanford University, Stanford, CA 94305, USA}
\affiliation{$^3$Stanford Institute for Materials and Energy Sciences, SLAC National Accelerator Laboratory and 
Stanford University, Menlo Park, CA 94025, USA}
\affiliation{$^4$ Department of Physics and Astronomy, University of Tennessee, Knoxville, TN 37996, USA}
\affiliation{$^5$ Joint Institute for Advanced Materials, The University of Tennessee, Knoxville, 425 Dougherty Engineering Building, Knoxville, TN 37996, USA}
\affiliation{$^6$Department of Physics and Astrophysics, University of North Dakota, Grand Forks, ND 58202, USA}

\date{\today}

\begin{abstract}
We present electron and phonon spectral functions calculated from determinant quantum Monte Carlo simulations of the half-filled two-dimensional Hubbard-Holstein model on a square lattice. By tuning the relative electron-electron ($e$-$e$) and electron-phonon ($e$-$ph$) interaction strengths, we show the electron spectral function evolving between antiferromagnetic insulating, metallic, and charge density wave (CDW) insulating phases. The phonon spectra concurrently gain a strong momentum dependence and soften in energy upon approaching the CDW phase. In particular, we study how the $e$-$e$ and $e$-$ph$ interactions renormalize the spectra, and find that the presence of both interactions suppresses the amount of renormalization at low energy, thus allowing the emergence of a metallic phase at intermediate coupling strengths. In addition, we find a modest enhancement of the $d$-wave pairing susceptibility in the metallic regime, although spin and charge correlations are still dominant at the temperatures considered in our study. These findings demonstrate the importance of considering the influence of multiple interactions in spectroscopically determining any one interaction strength in strongly correlated materials.
\end{abstract}

\pacs{71.10.Fd, 71.30.+h, 71.38.-k, 71.45.Lr, 74.72.-h} \maketitle

\section{Introduction}
Novel physics often emerges at the interface between competing ordered phases in condensed matter systems. This paradigm is universal across diverse classes of materials, including the colossal magnetoresistive effect near a phase boundary between charge ordering and ferromagnetism in the manganites~\cite{tokura}, the magnetoelectric effect occurring at a phase boundary in multiferroics~\cite{tokura2}, and a variety of emergent states at oxide interfaces.~\cite{hwang} In regions of phase competition, small changes to external parameters, such as doping, temperature, or pressure, can lead to large changes in materials properties. This sensitivity arises out of multiple interactions that exist on similar energy scales in strongly correlated materials. From a theoretical perspective, studying phase competition in a system with many strong interactions is a  challenge which requires non-perturbative methods.

The half-filled Hubbard-Holstein (HH) model, which includes both strong electron-electron ($e$-$e$) and electron-phonon ($e$-$ph$) interactions, provides a context to study phase competition in a model system. 
On a half-filled two dimensional square lattice, the Hubbard and Holstein models have instabilities towards $(\pi/a, \pi/a)$ antiferromagnetic (AFM) and charge density wave (CDW) orders, respectively.  The study of the HH model, in particular, also is  motivated by experimental evidence for strong $e$-$ph$ interactions, in addition to strong electronic correlations, in a variety of systems, including the high temperature superconducting cuprates~\cite{lanzara,shen_polaron,gunnarsson,cuk}, the  manganites,~\cite{millis,millis_polaron,mannella} the fullerenes,~\cite{gunnarsson2,durand,capone} and the rare-earth nickelates.~\cite{medarde,park,lau,johnston_nickelate,lee2012phase,kung2013,chuang2013}

Previous studies of ordered phases in the HH model have employed a variety of approaches. In one dimension, the HH phase diagram has  an intermediate metallic state between insulating AFM and CDW phases.~\cite{clay,hardikar,takada,fehske} In two dimensions, determinant quantum Monte Carlo (DQMC) studies~\cite{Nowadnick,Johnston} find evidence for an intervening metallic phase between the AFM and CDW states; the two dimensional phase diagram also has been studied with perturbative~\cite{berger,kumar} and strong coupling approaches.~\cite{hotta} In addition, both single-site dynamical mean-fleld theory (DMFT) and finite-size cluster DMFT  studies have been performed with mixed results.~\cite{bauer,bauer2,koller,koller2,sangiovanni,sangiovanni2,werner,macridin,khatami,murakami,murakami2} At zero temperature, no evidence for an intervening metallic phase was found with DMFT,~\cite{bauer} while finite temperature DMFT studies did find such a phase.~\cite{murakami} 

In this paper we  analytically continue DQMC calculations to obtain real frequency electron and phonon spectral functions for the half-filled HH model. The aim  is twofold: first, we provide additional evidence for and analysis of an intervening metallic phase, which we previously studied~\cite{Nowadnick,Johnston} using imaginary time quantities directly accessible from DQMC simulations.  We find that the spectral gap closes at the same $e$-$e$ and $e$-$ph$ interaction strengths where the metallic phase was identified from imaginary time quantities.
Second, we determine how the interplay of the $e$-$e$ and $e$-$ph$ interactions influences the observed renormalizations of the electron and phonon spectral functions.  

Both $e$-$e$ and $e$-$ph$ interactions  renormalize the electron spectral function relative to the bare non-interacting bandstructure: $e$-$ph$ coupling is revealed most commonly as ``kinks" in the low energy dispersion, while $e$-$e$ interactions can renormalize the spectra over a wide energy range. With increasing $e$-$ph$ coupling, the phonons gain a momentum-dependent dispersion, strongly renormalizing the phonon spectral function across all momenta as the system transitions to the CDW phase. Since both interactions are generally important in real materials, it is important to untangle their effects on the spectra. 

 In Sec.~\ref{methods} we present the HH model and our numerical methods. Sec.~\ref{spfn} analyzes the evolution of the electron spectral function with coupling strength,  Sec.~\ref{susc} focuses on the temperature- and phonon frequency- dependence of the AFM and CDW insulating states, and  Sec. ~\ref{phonon_spfn}  discusses the phonon spectral function. In Sec.~\ref{frequency} we study the temperature- and phonon phonon frequency- dependence of the intervening metallic phase, and in Sec.~\ref{sc} we examine the superconducting pairing susceptibilities. Finally, in Sec.~\ref{conclusion}, we summarize our results.

\section{Model and Methods \label{methods}}
The Hamiltonian for the two dimensional single-band HH model is $H = H_{\mathrm{kin}}+H_{\mathrm{lat}}+H_{\mathrm{int}}$, where
\begin{eqnarray}
H_{\mathrm{kin}}&=& -t \sum_{<ij>\sigma} c_{i\sigma}^\dagger c_{j\sigma}^{\phantom{\dagger}} -\mu \sum_{i\sigma}\hat{n}_{i\sigma} \\ \nonumber
H_{\mathrm{lat}}&=& \sum_i \Big ( \frac{M\Omega^2}{2}\hat{X}_i^2+\frac{1}{2M}\hat{P}_i^2 \Big ) \\
H_{\mathrm{int}}&=&U\sum_i \Big (\hat{n}_{i\uparrow}-\frac{1}{2} \Big )\Big  (\hat{n}_{i\downarrow}-\frac{1}{2} \Big ) -g\sum_{i\sigma}\hat{n}_{i\sigma}\hat{X}_i. \nonumber
\end{eqnarray}
Here $<$...$>$ denotes a sum over nearest neighbors, $c_{i\sigma}^{{\dagger}}$ creates an electron with spin $\sigma$ at site $i$, $\hat{n}_{i\sigma}=c_{i\sigma}^{{\dagger}} c_{i\sigma}^{\phantom{\dagger}}$, $\hat{X}_i$ and $\hat{P}_i$ are the atomic displacement and momentum operators at site $i$, $t$ is the nearest neighbor hopping, $\Omega$ is the phonon frequency, $U$ is the $e$-$e$ interaction strength, $g$ is the $e$-$ph$ interaction strength, and $\mu$ is the chemical potential  The dimensionless $e$-$ph$ coupling constant is defined as $\lambda=g^2/M\Omega^2 W$, where $W=8t$ is the electronic bandwidth. To account for the nonzero equilibrium lattice displacement present in the HH model, we set $\mu = -W\lambda$ to maintain half-filling at all coupling strengths.~\cite{Johnston} Throughout we take $t=1$, $M=1$,  and $a=1$ as our units of energy, mass, and length, respectively. 

The physics of the HH model is often analyzed using an effective-$U$ Hubbard model. By integrating out the phonons in a path integral framework, the HH model is mapped onto a Hubbard model with a frequency dependent effective interaction strength:
\begin{equation}
\label{eq:ueff}
U_{\mathrm{eff}}(\omega) = U-\frac{g^2}{M(\Omega^2-\omega^2)}.
\end{equation}
In the antiadiabatic limit ($\Omega \rightarrow \infty$), the  effective-$U$ becomes frequency independent: $U_{\mathrm{eff}} = U - \lambda W$, and the HH model maps to a static Hubbard model. For large $\Omega$ (and/or low energies $\omega$), the physics of the HH model approaches that of a $U_{\mathrm{eff}}$ Hubbard model. For small phonon frequencies $\Omega$, retardation effects are important and the physics of these two models differs substantially.~\cite{bauer,Nowadnick}

We simulate the HH model on a square lattice at half-filling using DQMC, which is a numerically exact method that treats the $e$-$e$ and $e$-$ph$ interactions on an equal footing and non-perturbatively.~\cite{BSS,White,Johnston} The presence of simultaneously non-zero $e$-$e$ and $e$-$ph$ couplings in the HH model introduces a fermion sign problem at half filling.~\cite{Johnston} DQMC requires an imaginary time discretization; we use a  step size  $\Delta\tau=0.125/t$ or smaller for all results shown in this paper.

The DQMC simulation provides the imaginary time electron and phonon Green's functions, $G({\bf K},\tau)= <T c_{\bf K}(\tau)c_{\bf K}^\dagger(0)>$ and $D({\bf K},\tau)=<T \hat{X}_{\bf K}(\tau) \hat{X}_{\bf K}(0)>$ on a discrete grid of momentum space points \{${\bf K}$\}, determined by the size of the simulation cluster with periodic boundary conditions.  The low energy electron and phonon spectral weights are directly accessible from the imaginary time Green's functions via the relations~\cite{trivedi}
\begin{equation}
\beta G({\bf K},\tau=\beta/2)=\frac{\beta}{2}\int d\omega \frac{A({\bf K},\omega)}{\cosh(\beta\omega/2)}
\label{eq:gbeta}
\end{equation}
and
\begin{equation}
\beta D({\bf K},\tau=\beta/2)=\frac{\beta}{2}\int d\omega \frac{\omega B({\bf K},\omega)}{\sinh(\beta\omega/2)}.
\end{equation}

Obtaining the full frequency-dependent electron and phonon spectral functions, $A({\bf K},\omega)$ and $B({\bf K},\omega)$, requires  numerical analytic continuation to real frequencies; in this work we utilize the Maximum Entropy method (MEM).~\cite{Jarrell} The MEM technique requires a model function for use in determining an entropic prior; for electronic spectral function analytic continuations we use an uninformative (``flat") model, while for phonon spectral function analytic continuations we use a Lorentzian model peaked at the bare phonon frequency $\Omega$ and of width $t$ for all momenta and $e$-$ph$ coupling strengths. We have checked that the spectral functions are robust against reasonable changes to these models.

For the high resolution electron spectral function plots shown in this paper, we employ the following interpolation scheme.
After analytic continuation, we obtain the electronic self-energy $\Sigma({\bf K},\omega)$ from Dyson's equation: $\Sigma({\bf K},\omega) = \omega - \epsilon_{\bf K} - G^{-1}({\bf K},\omega)$. For a cluster of a sufficient size, the momentum dependence of the self-energy is well approximated by a linear interpolation onto a finely-spaced momentum mesh  ${\bf k}$: $\Sigma({\bf K},\omega) \rightarrow \Sigma({\bf k},\omega)$.~\cite{moritz2} The interpolated electronic Green's function $G({\bf k},\omega)$ on this fine momentum mesh is obtained by another application of Dyson's equation: $G({\bf k},\omega)=[\omega-\epsilon_{\bf k}-\Sigma({\bf k},\omega)]^{-1}$. An analysis of the dependence of the spectra on cluster size and the robustness of the interpolation method is presented in  Appendix A.

\section{Electron spectral function \label{spfn}}
\begin{figure*}
 \begin{center}
\includegraphics[width=\textwidth]{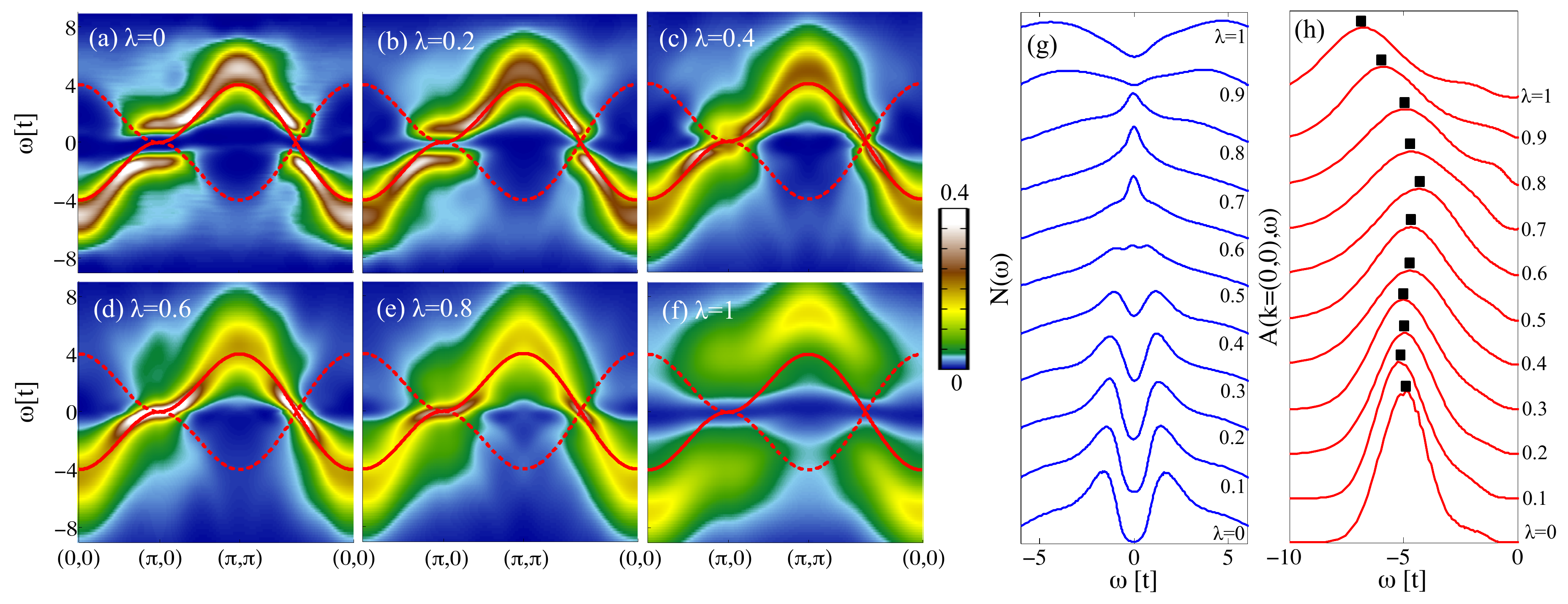}
\caption{\label{fig:1}  (a)-(f) Spectral functions $A({\bf k},\omega)$ along high symmetry cuts through the Brillouin zone for various $e$-$ph$ interaction strengths $\lambda$. The solid red line indicates the non-interacting tight binding band structure, and the dashed red line indicates the  band shifted by ($\pi$, $\pi$). (g) density of states $N(\omega)$ and (h) $A({\bf k}=(0,0),\omega)$ for increasing $\lambda$ (bottom to top). The black squares in (h) denote the maxima of the spectra. Note that the spectra in (g) and (h) are artificially offset for clarity. The remaining simulation parameters are $U=6t$, $\Omega=t$, $\beta=4/t$, and $N=8\times8$.}
\end{center}
\end{figure*}

The electronic spectral function $A({\bf k},\omega)$ is shown in Fig.~\ref{fig:1} for $U$=$6t$ and several values of the $e$-$ph$ coupling strength $\lambda$ along high symmetry cuts in the Brillouin zone.  When $\lambda$=0 (Fig.~\ref{fig:1}(a)), spectral weight is concentrated in the lower and upper Hubbard bands (LHB and UHB) centered at (0,0) and ($\pi$,$\pi$), respectively. Tails of spectral weight disperse towards the Fermi level, which are precursors to the quasiparticle band that develops in the doped Hubbard model.~\cite{moritz2} A well defined Mott gap is clearly visible in these momentum space cuts at ${\bf k}=(\pi,0)$ and $(\pi/2,\pi/2)$. These results agree well with previous Hubbard model spectral function studies in two dimensions.~\cite{preuss,grober,moritz,moritz2}

As $\lambda$ increases, the spectral function evolves from that of a Mott insulator (Fig.~\ref{fig:1}(a-c)) to a metallic system (Fig.~\ref{fig:1}(d-e)), and finally to a CDW insulator (Fig.~\ref{fig:1}(f)). The closing and reopening of the spectral gap with increasing $\lambda$ is also clear from the density of states (DOS), defined as $N(\omega)=\frac{1}{N}\sum_{\bf K} A({\bf K},\omega)$  and shown in Fig.~\ref{fig:1}(g) (note that the sum is performed over the allowed ${\bf K}$ in the $8 \times 8$ cluster). In addition, the spectra change from being dominated by $e$-$e$ interaction effects to displaying a mixture of properties from the $e$-$e$ and $e$-$ph$ interactions, although the interplay of interactions affects the low and high energy spectral properties differently.

Once $\lambda$ becomes nonzero (Fig.~\ref{fig:1}(b-c)), the Mott gap clearly narrows, due to the reduction in the effective $U$ at low energy by the $e$-$ph$ interaction. Note that for $\omega >> \Omega$, the Hubbard $U$ is essentially unrenormalized in  Eq.~(\ref{eq:ueff}) by the $e$-$ph$ interaction. As a result, the LHB and UHB peak positions remain unchanged from the Hubbard model results. Fig.~\ref{fig:1}(h) shows the evolution of $A({\bf k}=(0,0),\omega)$ with $\lambda$, from which it is clear that the LHB peak location remains essentially fixed up to a relatively large $\lambda$, however, the spectra broaden due to the presence of additional scattering from phonons. 

Between $\lambda$=0.4 and 0.5, the Mott gap closes completely, and a quasiparticle peak appears at the Fermi level as the system transitions to a metallic state that persists over an intermediate range of $\lambda$, shown here for $\lambda$=0.6 and 0.8 (Fig.~\ref{fig:1}(d-e)). 
There is a clear ``kink" in the band dispersion at $\omega\approx 1-2t$ in Fig.~\ref{fig:1}(d-e), which separates the sharp, weakly dispersing (relative to the bare band) low energy quasiparticle band from the broad high energy spectra. The peak position of the LHB/UHB softens slightly, while the width increases noticeably  in Fig.~\ref{fig:1}(h). The coupling strengths $\lambda$ at which the crossovers between the metallic and insulating phases occur agree well with our determination of these crossovers from imaginary time quantities, as presented in Ref.~\onlinecite{Nowadnick}.

What is the origin of the kink feature in these spectra? In a metallic system of coupled electrons and phonons with a weak to intermediate interaction strength, a kink occurs at the phonon frequency, with significant spectral broadening for $\omega>\Omega$ due to the onset of the imaginary part of the self-energy.~\cite{Mahan} The energy scale of the kinks in Fig.~\ref{fig:1}(d-e) is similar to that of the bare phonon frequency $\Omega=t$, however, the renormalized phonon frequency, which should be the relevant energy scale for setting an $e$-$ph$ kink, is substantially lower, as will be discussed in Section~\ref{phonon_spfn}. In addition, $e$-$e$ interactions also produce band renormalizations; in the doped Hubbard model a kink-like ``waterfall" feature occurs at $\omega\approx 1.5-2t$ at the crossover between the shallow quasiparticle band and the LHB/UHB. Because in the present simulation  the phonons and $e$-$e$ band renormalizations are of a similar energy scale, we conclude that the kinks in Fig.~\ref{fig:1}(d-e) have a mixed origin from both the $e$-$e$ and $e$-$ph$ interactions. ~\cite{foot1}
 
Finally, as $\lambda$ increases further, the quasiparticle peak disappears, and a new gap opens at the Fermi level (Fig.~\ref{fig:1}(f)), this time originating from the CDW insulating state. The peak position of the high energy spectral weight moves to substantially higher energies (Fig.~\ref{fig:1}(h)), as it now contains contributions from the UHB/LHB and the phonon sidebands which appear in the Holstein model in the CDW state.~\cite{marsiglio,vekic,niyaz} Finally, in both Fig.~\ref{fig:1}(a) and (f), a faint trail of spectral weight follows the folded bands (bands displaced by ($\pi$, $\pi$), shown with dashed lines), which are expected in the ordered insulating states. 

\begin{figure*}
 \begin{center}
\includegraphics[width=\textwidth]{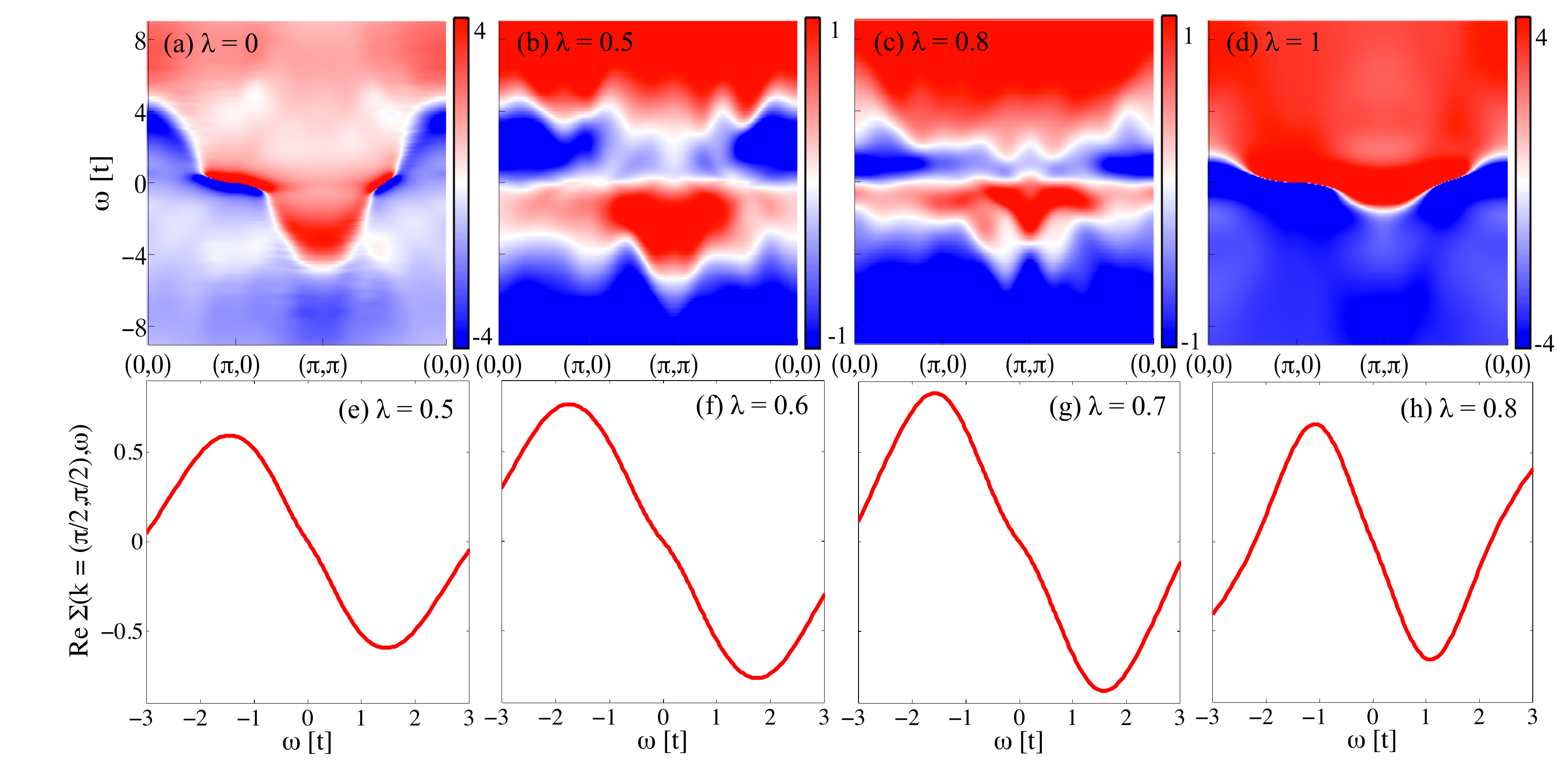}
\caption{\label{fig:2} Real part of the self-energy $\Sigma'({\bf k},\omega)$ along high symmetry cuts through the Brillouin zone for (a) $\lambda=0$, (b) $\lambda$=0.5,  (c) $\lambda$=0.8, and (d) $\lambda$=1. (e-h) Real part of the self-energy at ${\bf k}=(\pi/2,\pi/2)$ for $\lambda$=0.5-0.8. The remaining simulation parameters are $U=6t$, $\beta=4/t$, $\Omega=t$, and $N=8\times8$.}
\end{center}
\end{figure*}

We further analyze this insulator - metal - insulator transition by considering the self-energy $\Sigma({\bf k},\omega)$ in Fig.~\ref{fig:2}. The real part of the self-energy, $\Sigma'({\bf k},\omega)$, for high symmetry cuts through the Brillouin zone, is shown in Fig.~\ref{fig:2}(a-d). In the Mott (Fig.~\ref{fig:2}(a)) and CDW (Fig.~\ref{fig:2}(d)) insulating phases, the self-energy diverges at the Fermi level, signaling the presence of a gap (visible here at ${\bf k}$ = $(\pi,0)$ and $(\pi/2,\pi/2)$). At other momenta, $\Sigma'$ evolves smoothly with energy. In the metallic phase, shown in Fig.~\ref{fig:2}(b-c) for $\lambda=0.5$ and 0.8, respectively, $\Sigma'$ has a negative slope as it passes through the Fermi level. 

Fig.~\ref{fig:2}(e-h) shows $\Sigma^\prime$ at $(\pi/2,\pi/2)$ for several values of $\lambda$ in the metallic regime, from which it is clear that the self-energy is Fermi-liquid like (Re $\Sigma$ $\propto$ $-\omega$).~\cite{foot2} The peak in $\Sigma^\prime$ occurs between $t$ and $2t$, which is larger than the phonon energy, suggesting that the electronic correlations have a strong influence on setting the kink energy scale. However, the peak location changes with $\lambda$, in particular, moving to lower energy by $\lambda=0.8$ in Fig.~\ref{fig:2}(h), which demonstrates that phonons also play an important role. Consistent with this picture, a recent DMFT study~\cite{bauer3} found that electronic correlations can push an $e$-$ph$ kink position above the phonon energy scale. Finally, it is important to note that the interpolation procedure used to create these spectra from a coarser momentum mesh set by the size of the real space cluster could influence the kink location, but we compared spectra from $N=8\times 8$, $N=10\times 10$, and $N=12\times 12$ clusters and found almost identical kink structures, so believe that this effect is minimal. 
 
 \begin{figure}
\includegraphics[width=0.3\textwidth]{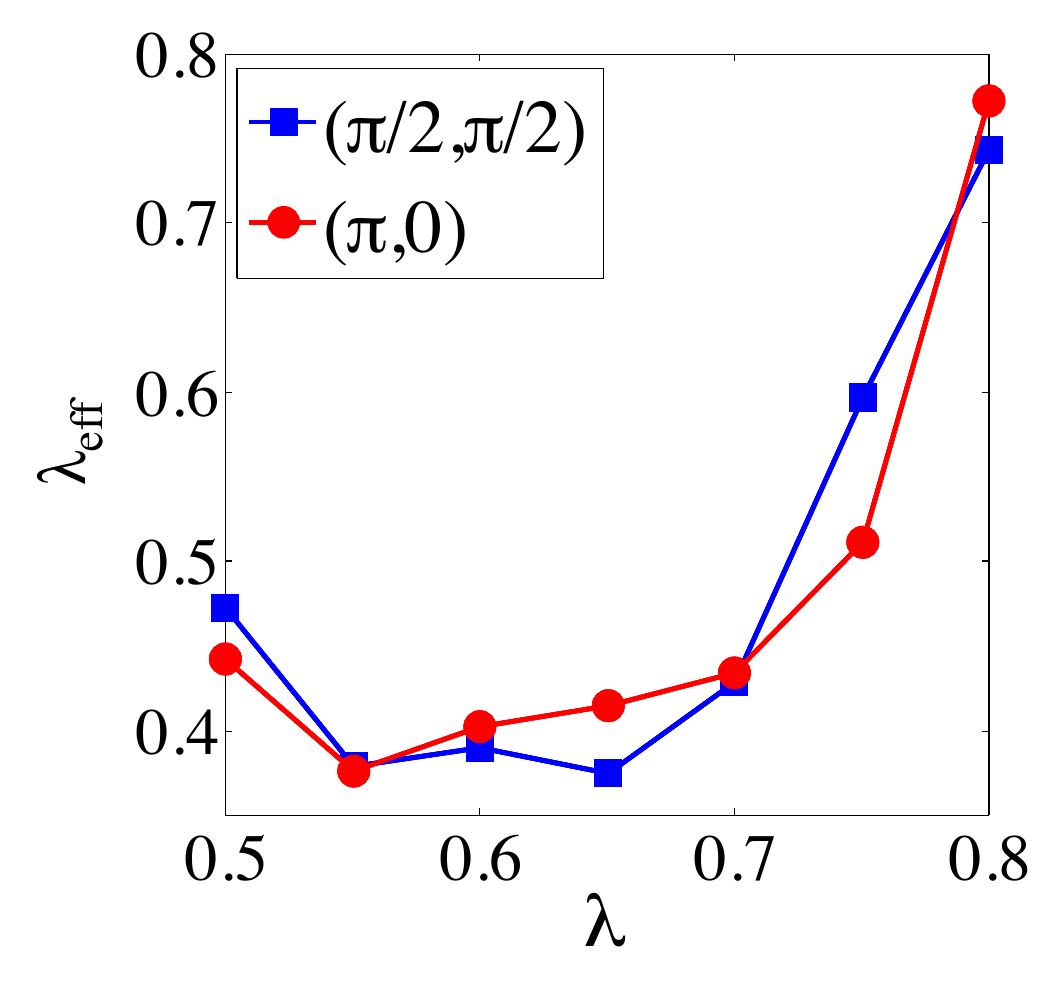}
\caption{\label{fig:8}  $\lambda_{\mathrm{eff}}$ as defined in Eq.~\ref{eq:lambda_star}, as a function of $e$-$ph$ coupling strength $\lambda$ with the $e$-$e$ interaction set to $U=6t$. Other simulation parameters are $\beta = 4/t$, $\Omega = t$, $N = 8\times 8$.}
\end{figure}
 
Since the system is Fermi liquid-like in the intermediate metallic regime, the strength of the band renormalization is given by the ratio of the renormalized and bare Fermi velocities  $v_F/v_F^{(0)}=1/(1+\lambda_{\mathrm{eff}})$, where
\begin{equation}
\lambda_{\mathrm{eff}}({\bf k}) = -\frac{\partial \Sigma '({\bf k},\omega)}{\partial \omega} \Big |_{\omega = 0}.
\label{eq:lambda_star}
\end{equation}
We emphasize that both the $e$-$e$ and the $e$-$ph$ interaction contribute to $\lambda_{\mathrm{eff}}$. 

Fig.~\ref{fig:8} show  $\lambda_{\mathrm{eff}}$  as a function of $e$-$ph$ interaction strength $\lambda$ and with a fixed $e$-$e$ interaction $U$=6$t$. Note that we can only compute $\lambda_{\mathrm{eff}}$ in the intermediate-$\lambda$ regime where $\Sigma'$ has a well defined slope at zero energy. We extract $\lambda_{\mathrm{eff}}$ at both $(\pi/2,\pi/2)$ and $(\pi,0)$, and find qualitatively similar behavior at both momentum points. If the band renormalization were only due to the $e$-$ph$ interaction, or if the $e$-$e$ and $e$-$ph$ interactions cooperated (so that the renormalizations from the two interactions were additive), we would expect that $\lambda_{\mathrm{eff}}$ would increase monotonically with $\lambda$ for fixed $U$.~\cite{iwasawa}  Instead, we find that $\lambda_{\mathrm{eff}}$ displays non-monotonic behavior. As $\lambda$ increases from 0.5 to 0.6, $\lambda_{\mathrm{eff}}$ in fact decreases, meaning that the effective correlation strength in the system has gone down as the $e$-$e$ and $e$-$ph$ interactions partially cancel each other. Then, between $\lambda$= ~0.6 and ~0.7, $\lambda_{\mathrm{eff}}$ essentially does not change, and only when $\lambda$ passes 0.7, $\lambda_{\mathrm{eff}}$ starts to increase rapidly.  The coupling at which  $\lambda_{\mathrm{eff}}$ begins to increase quickly approximately corresponds to  that where U$_{\mathrm{eff}}$=0 ($\lambda=0.75$). This suppression of the correlation strength $\lambda_{\mathrm{eff}}$ occurs in the parameter regime where the spectral functions exhibit metallic behavior in Fig.~\ref{fig:1}. 

To summarize this section, our analysis of the spectral function evolution with $\lambda$ revealed that  the low and high energy spectral features respond to the simultaneous presence of $e$-$e$ and $e$-$ph$ interactions in different ways. The high energy part of the spectra become increasingly incoherent as $\lambda$ increases due to presence of scattering from both interactions, and are broader than spectra with only one of $U$ and $\lambda$ nonzero. In contrast,  there are fairly well-defined coherent quasiparticles, dressed in a Fermi-liquid manner by phonon and Coulomb interactions, present at low energies. This occurs due to the presence of both interactions, since for $U=6t$, $\lambda\approx$ 0.6 the system would be either Mott or CDW insulating if only one interaction were present. Thus, the system transfers spectral weight back to the Fermi level in order to reform quasiparticles as $\lambda$ is increased. It is natural to ask whether other phases besides metallicity can appear, such as superconductivity, which we investigate in Sec.~\ref{sc}.

\section{Temperature- and $\Omega$-dependence of insulating states \label{susc}}

\begin{figure*}
 \begin{center}
\includegraphics[width=0.8\textwidth]{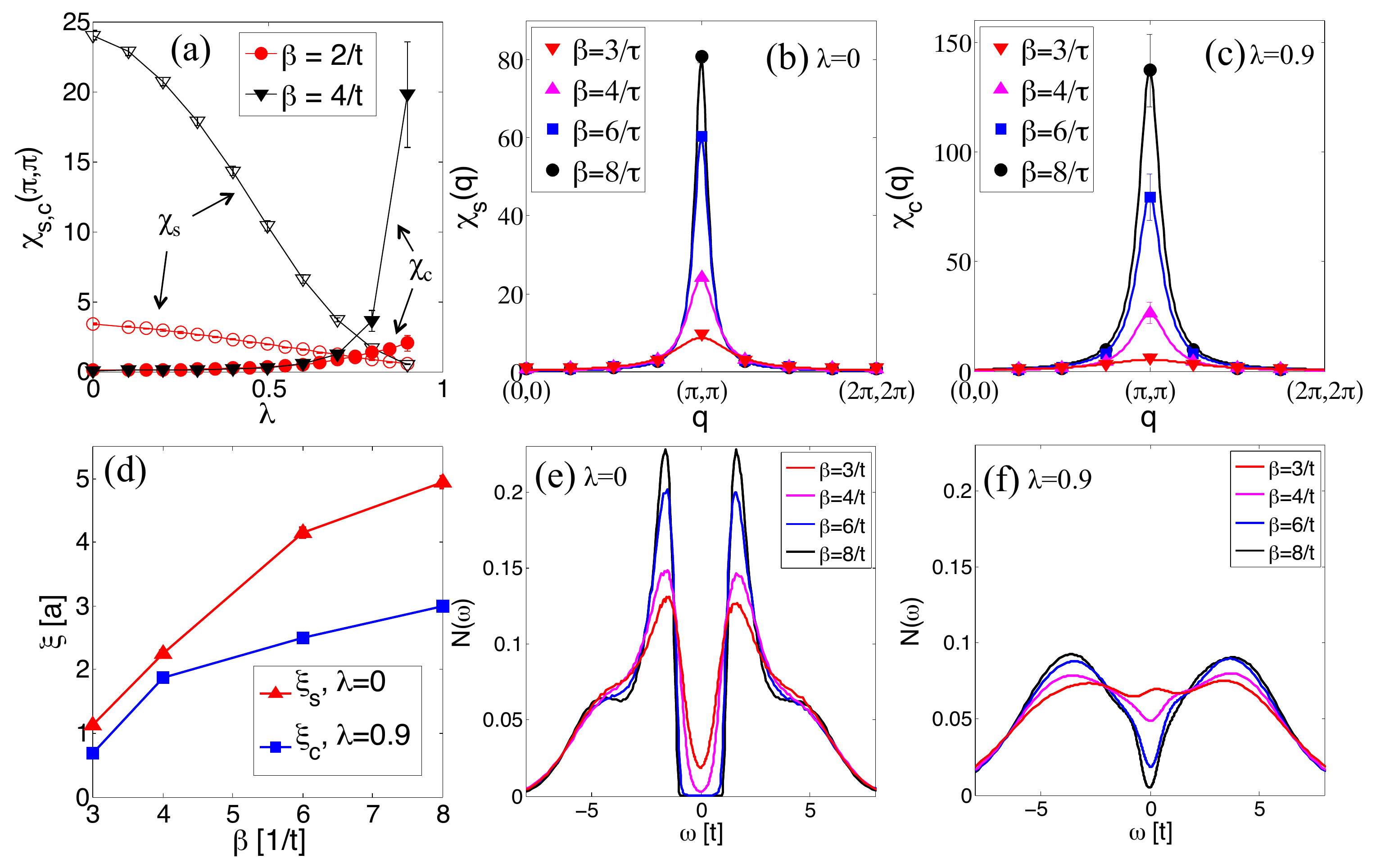}
\caption{\label{fig:11} Temperature dependence of the spin and charge susceptibilities $\chi_s({\bf q})$ and $\chi_c({\bf q})$ for $U=6t$ evaluated in several ways. (a) $\chi_s(\pi,\pi)$ (open symbols) and $\chi_c(\pi,\pi)$ (solid symbols)  as a function of $\lambda$. (b) $\chi_s({\bf q})$ for $\lambda = 0$,  and (c) $\chi_c({\bf q})$ for  $\lambda=0.9$ for several temperatures. The points are the DQMC data, while the lines are Lorentzian fits.  (d) coherence length $\xi$ of $\chi_s({\bf q})$ and $\chi_c({\bf q})$ from (b) and (c) as a function of temperature. DOS $N(\omega)$ for (e) $\lambda=0$, and (f) $\lambda=0.9$. Other simulation parameters are $\Omega = t$ and $N = 8\times 8$.}
\end{center}
\end{figure*}

Given the complex evolution of the spectra with interaction strength discussed in the previous section, it is interesting to check how other parameters, such as temperature and phonon frequency, influence these results. In this section we focus on the AFM and CDW insulating states, while in Sec.~\ref{frequency} we study the metallic state.
We first examine the  spin and charge susceptibilities, defined as:
\begin{equation}
\chi_{s,c}({\bf q})=\frac{1}{N} \int_0^\beta d\tau <T_\tau \hat{O}^{\phantom{\dagger}}_{s,c}({\bf q},\tau) \hat{O}^\dagger_{s,c}({\bf q},0)>
\end{equation}
where $\hat{O}_s({\bf q})=\sum_i e^{i{\bf q}\cdot {\bf R}_i}(\hat{n}_{i\uparrow}-\hat{n}_{i\downarrow})$, and $\hat{O}_c({\bf q})=\sum_{i} e^{i{\bf q}\cdot {\bf R}_i}(\hat{n}_{i\uparrow}+\hat{n}_{i\downarrow})$.

Fig.~\ref{fig:11}(a) shows $\chi_s(\pi, \pi)$ and $\chi_c(\pi, \pi)$ as  functions of $\lambda$ at two different temperatures. As the temperature is lowered, $\chi_s(\pi,\pi)$ and $\chi_c(\pi,\pi)$ grow quickly for small and large $\lambda$, respectively. These susceptibilities show a weaker temperature dependence in the intermediate $\lambda$ range, where the spectral functions exhibit metallic behavior in Fig.~\ref{fig:1}. 

Fig.~\ref{fig:11}(b) and (c) show the temperature dependence of $\chi_s({\bf q})$ and  $\chi_c({\bf q})$ in parameter regimes where spin and charge correlations are large, respectively. 
The susceptibilities $\chi_{s,c}({\bf q})$ are fit to two dimensional Lorentzian functions,
$L({\bf q}) = A/[(q_x - B)^2 + (q_y - C)^3 +(1/\xi)]^2$
which are shown as solid lines in Fig.~\ref{fig:11}(b-c) along the $(0,0)\rightarrow (\pi,\pi)$ cut.  The correlation lengths $\xi$ extracted from these fits are shown in Fig.~\ref{fig:11}(d). Note that in two dimensions, the transition to long range AFM spin order occurs at $T$=0 due to the Mermin-Wagner theorem, while the CDW order has a finite $T$ transition. Interestingly, while $\chi_c(\pi,\pi)$ grows faster with lowering temperature than $\chi_s(\pi,\pi)$ (compare peak heights in Fig.~\ref{fig:11}(b-c)), the correlation length $\xi_c$ grows more slowly than $\xi_s$.  
Fig.~\ref{fig:11}(e) and (f) show the temperature dependence of the DOS $N(\omega)$ for the same parameter sets as the susceptibilities shown in (b) and (c). As temperature lowers, the Mott gap (e) and CDW gap (f) develop. As the Mott gap opens, spectral weight is shifted into coherence peaks that sharpen with lowering temperature, while the opening of the CDW gap is characterized by a suppression of spectral weight over a wide energy range. 

\begin{figure}
\includegraphics[width=0.35\textwidth]{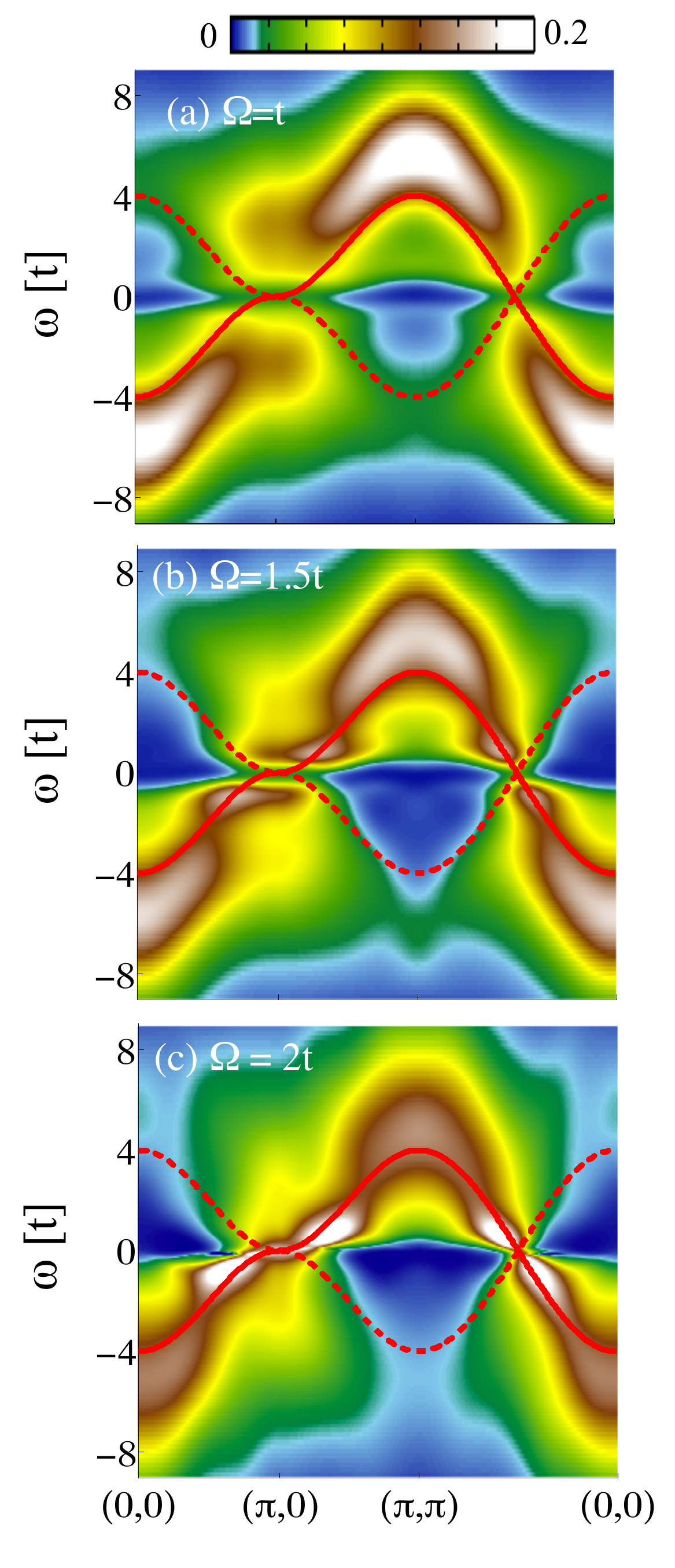}
\caption{\label{fig:5} Spectral functions $A({\bf k},\omega)$ for high-symmetry cuts through the Brillouin zone for (a) $\Omega = t$, (b) $\Omega = 1.5t$, and (c) $\Omega = 2t$. The remaining simulation parameters are $U=6t$, $\lambda = 0.9$, $\beta = 6/t$, and $N = 8\times 8$.}
\end{figure}

Since the phonon frequency influences the interaction strength in Eq.~\ref{eq:ueff}, it also will influence the coupling strengths at which the transitions between metallic and insulating states occurs. To demonstrate this effect, 
 Fig.~\ref{fig:5} shows  $A({\bf k},\omega)$ at fixed  $U$=6$t$ and $\lambda$=0.9 for three different phonon frequencies. When $\Omega=t$ (Fig.~\ref{fig:5}(a)), there is a clear CDW gap, as well as spectral weight following the folded bands. Upon increasing the phonon frequency to $\Omega=1.5t$ (Fig.~\ref{fig:5}(b)), the size of the gap decreases, as does the intensity of the spectral weight tracking the folded bands. Finally, when $\Omega=2t$ (Fig.~\ref{fig:5}(c)), the CDW gap has completely closed. In Fig.~\ref{fig:5}(b) and (c), there are clear kinks in the band dispersion which move to higher energy as the phonon frequency increases from (b) to (c), thus showing that this kink energy scale is clearly sensitive to the phonon frequency. Because  the phonon frequency impacts the extent of the CDW phase, it may be added as a third axis to the HH phase diagram at half filling, as discussed previously in the context of the one dimensional HH model.~\cite{hardikar} 

\section{Phonon spectral function \label{phonon_spfn}}
\begin{figure*}
 \begin{center}
\includegraphics[width=0.9\textwidth]{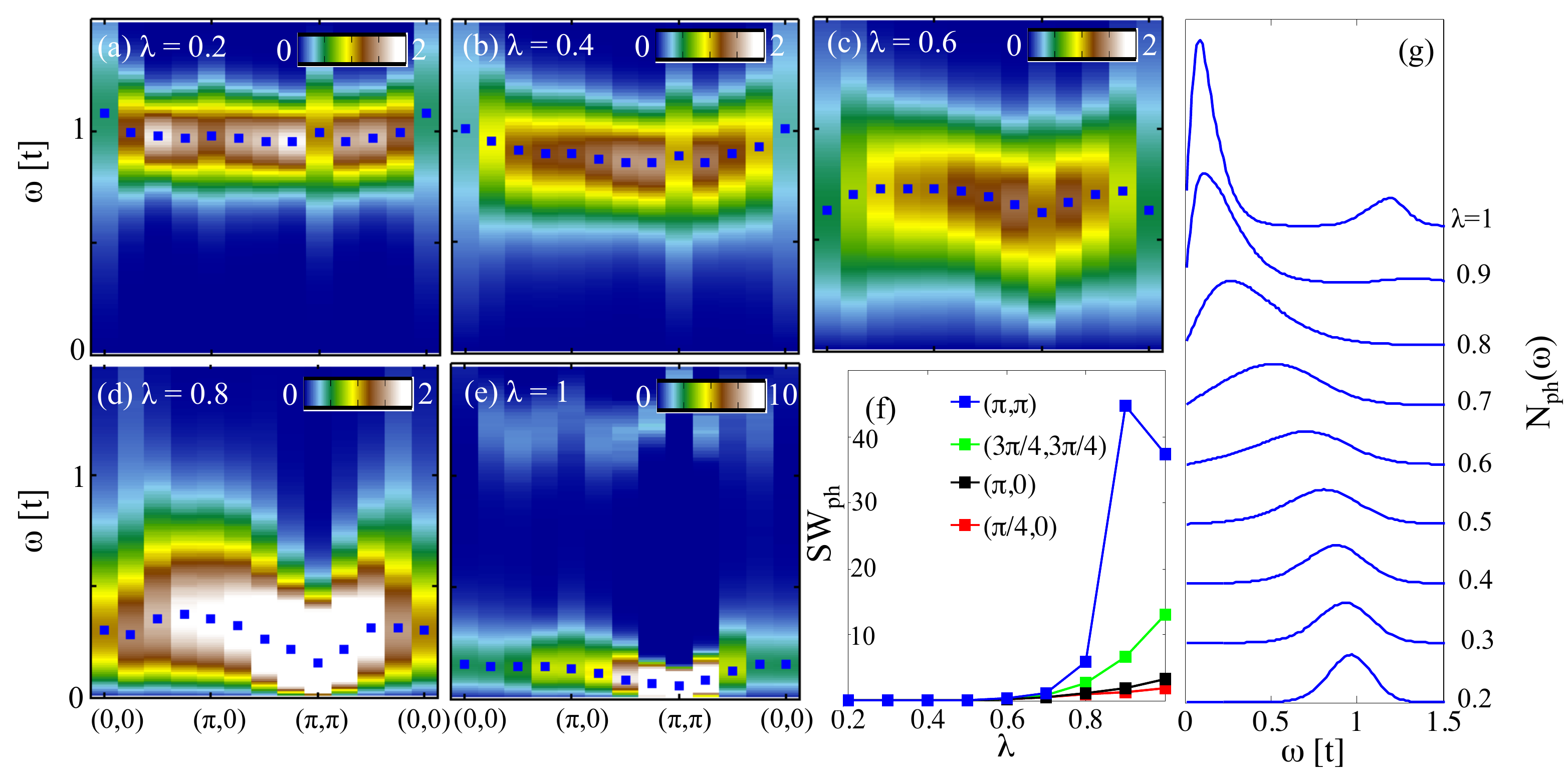}
\caption{\label{fig:3} (a-e) Phonon spectral function $B({\bf q},\omega)$ along high symmetry Brillouin zone cuts for various $e$-$ph$ strengths $\lambda$. The maximum at each ${\bf q}$ is indicated by blue squares. (f) The integrated phonon spectral weight as a function of $\lambda$ for several momentum points. (g) the phonon density of states $N_{ph}(\omega)$. The remaining simulation parameters are $U=6t$, $\beta=4/t$, $\Omega=t$, and $N=8\times8$.}
\end{center}
\end{figure*}

We now consider the phonon spectral function $B({\bf q},\omega)$. Fig.~\ref{fig:3}(a-e) shows $B({\bf q},\omega)$ along high symmetry Brillouin zone cuts for $U$=6$t$ and several $e$-$ph$ coupling strengths $\lambda$, while Fig.~\ref{fig:3}(g) shows the evolution of the phonon DOS, $N_{ph}(\omega)=\frac{1}{N}\sum_{\bf q} B({\bf q},\omega)$, with $\lambda$ (the sum over ${\bf q}$ encompasses the allowed momenta in the $N=8\times 8$ cluster). Note that when $\lambda=0$, the phonon spectral function is momentum independent: $B({\bf q},\omega) = \delta(\omega -\Omega)-\delta(\omega+\Omega)$. For $\lambda$=0.2-0.4 (Fig.~\ref{fig:3}(a-b)), $B({\bf q},\omega)$  remains fairly momentum-independent with the peak  at $\approx\Omega$ for all momenta, although the peak width grows as $\lambda$ increases from 0.2 to 0.4 due to increasing $e$-$ph$ scattering. As $\lambda$ increases to 0.6-0.8 (Fig.~\ref{fig:3}(c-d)), the peak in the phonon spectral function moves to lower energy and the width increases at all momenta. In addition, the phonons gain a  momentum-dependent dispersion  $\omega_{\bf q}$, with the spectral function peak at the  CDW ordering vector ${\bf Q}_{CDW}$=($\pi$, $\pi$) softening to the lowest energy. 
 At $\lambda$=1 (Fig.~\ref{fig:3}(e)), when the CDW gap is open in the spectral function in Fig.~\ref{fig:1}(f), the  peak at $(\pi,\pi)$ occurs very near zero frequency. A momentum-dependent phonon dispersion that goes to zero frequency at ${\bf Q}_{CDW}$ has previously been reported for the two dimensional Holstein model.~\cite{vekic}

The momentum-dependence of the phonon dispersion can be obtained by writing the Dyson's equation for the phonon Green's function:~\cite{Mahan}
\begin{equation}
D({\bf q},\omega) = \frac{2\Omega}{\omega^2 - \Omega^2 + 2g^2\Omega \chi_c({\bf q},\omega)}
\end{equation}
Then the renormalized phonon frequency $\omega_{\bf q}$ is given by the pole of this equation:
\begin{equation}
\omega_{\bf q}^2 = \Omega^2(1-2g^2\chi_c^\prime({\bf q},\omega_{\bf q})/\Omega)
\end{equation}
where $\chi_c^\prime({\bf q},\omega) $ is the real part of the charge susceptibility. When a CDW forms, the phonon  frequency falls to zero at the CDW ordering vector ${\bf Q}_{CDW}$. There are two possible mechanisms for this:  either $\chi_c^\prime({\bf q},\omega_{\bf q})$ grows large, or the $e$-$ph$ coupling $g$ becomes strong. The first case is the Peierls picture, where the Fermi surface is strongly nested at a particular ${\bf Q}_{CDW}$ and thus creating a CDW for arbitrarily weak coupling ($g\approx 0$). In the second case, driven by strong coupling, the Fermi surface plays a less important role.  In Fig.~\ref{fig:3}(c-e) the phonon dispersion is strongly renormalized at all momenta, indicating  the importance of strong $e$-$ph$ coupling in forming the CDW phase in the HH model. Note that here $\Omega=t$; lower phonon frequencies may lead to a more Peierls-like CDW. 



The renormalized phonon frequency, integrated over momenta, can be estimated from the peak positions in $N_{ph}(\omega)$ in Fig.~\ref{fig:3}(g), and it moves to very low energy as the coupling increases.  While this discussion has so far focused on the evolution of the phonon spectral function with $\lambda$, $U$ also plays an important role in delaying the softening of the phonon spectra to relatively large $\lambda$. Here, the spectra do not become significantly renormalized until $\lambda\approx 0.6$, while in the Holstein model a CDW forms at weak coupling.

The phonon spectral function, being bosonic, has normalization condition
\begin{equation}
\int_{-\infty}^\infty \frac{d\omega}{2\pi}  \frac{B({\bf q},\omega)}{\omega} = \int_0^\beta d\tau D({\bf q},\tau)
\end{equation}
 Therefore, as $\lambda$ increases, the phonon spectral weight increases at all momenta in Fig.~\ref{fig:3}(a-e), although most substantially at ${\bf Q}_{CDW}$=($\pi$, $\pi$).  The phonon spectral weight at wavevector ${\bf q}$ and temperature $1/\beta$ is given by 
\begin{equation}
SW_{ph} ({\bf q}) = \int_{-\infty}^\infty \frac{d\omega}{2\pi} n_B(\omega) B({\bf q},\omega)
\end{equation}
where $n_B(\omega)$ is the Bose factor. We plot $SW_{ph}({\bf q})$   as a function of $\lambda$ in Fig.~\ref{fig:3}(f) for several representative ${\bf q}$. As $\lambda$ increases and the system moves towards the CDW state, $SW_{ph}$ increases by orders of magnitude at $(\pi,\pi)$. The increase in spectral weight at momenta near $(\pi, \pi)$ is also substantial (($3\pi/4, 3\pi/4)$ is shown here), while there is a moderate increase even for momenta far away from $(\pi,\pi)$ (see $(\pi,0)$ and $(\pi/4,0)$). Therefore, the transition to the CDW phase involves a significant softening and increasing occupation of phonon modes at all momenta across the Brillouin zone. The increased phonon occupations across many momenta is likely due to phonon scattering events at these wavevectors.

Finally, we note that a second, higher energy peak appears in $N_{ph}(\omega)$ for $\lambda=0.9-1$ (Fig.~\ref{fig:3}(g)), this high energy spectral weight is also visible in Fig.~\ref{fig:3}(e). We have tried a variety of model functions  in the MEM analytic continuation, and find that this peak robustly appears for many different models. 
A DMFT study of the HH model~\cite{koller2} found that the phonon spectral function gains a two-peak structure as the system transitions to a bipolaronic CDW state; the two-peak structure in our phonon spectral functions likely has the same origin.

\section{Temperature- and $\Omega$-dependence of metallic state \label{frequency}}

\begin{figure}
\includegraphics[width=0.35\textwidth]{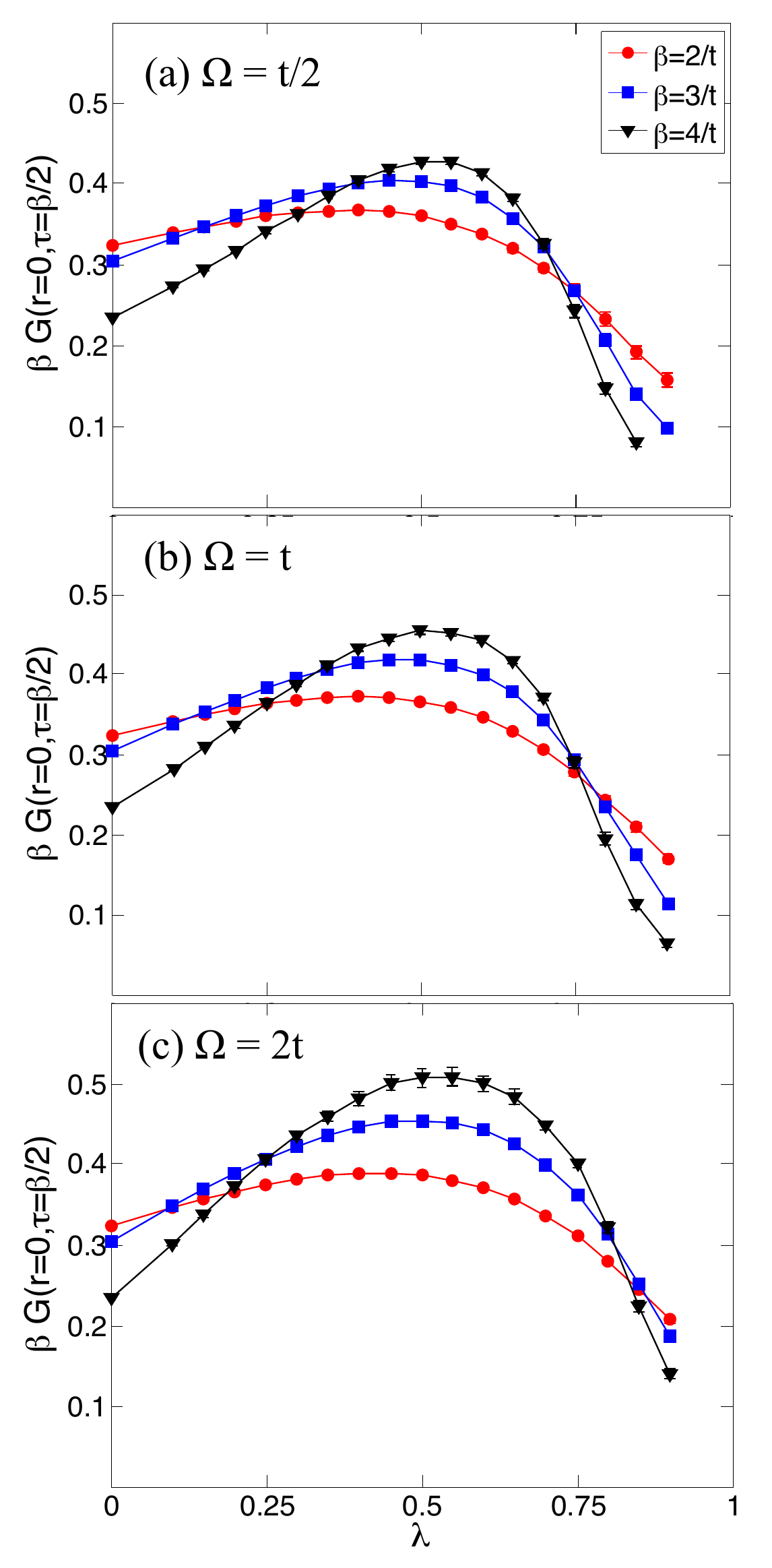}
\caption{\label{fig:7} Low energy spectral weight $\beta G({\bf r}=0,\tau=\beta/2)$ as a function of $\lambda$ at various temperatures for phonon frequencies $\Omega=$ (a) $t/2$, (b) $t$, and (c) $2t$. The remaining simulation parameters are $U=5t$ and $N=8\times 8$.}
\end{figure}

We now study the influence of the phonon frequency $\Omega$ on the metallic phase. Fig.~\ref{fig:7} shows the low energy  spectral weight $\beta G({\bf r}=0,\tau=\beta/2)=\sum_{\bf K}\beta G({\bf K},\tau=\beta/2)$ (abbreviated from here on as $\beta G_{\beta/2}$), as defined in Eq.~\ref{eq:gbeta} for fixed $U$ and $\lambda$ and three different phonon frequencies. In the low temperature limit, $\beta G_{\beta/2}$ is 0 if an insulating gap is present, and finite if a band disperses through the Fermi level. Outside of this limit, the temperature dependence of $\beta G_{\beta/2}$ yields information about insulating versus metallic behavior: with lowering temperature, the magnitude of $\beta G_{\beta/2}$ falls if an insulating gap is opening, while it increases in a metallic state as the quasiparticle peak sharpens. 

At all phonon frequencies considered in Fig.~\ref{fig:7}, at small and large $\lambda$, the magnitude of $\beta G_{\beta/2}$ decreases as the temperature is lowered, as the Mott and CDW gaps open, respectively. However, at intermediate $\lambda$, $\beta G_{\beta/2}$ increases as the temperature is lowered, which is indicative of a metallic state intervening between the Mott and CDW insulating states. The range of $\lambda$ over which $\beta G_{\beta/2}$ increases with decreasing temperature grows  with increasing $\Omega$: for $\Omega=t/2$ (Fig.~\ref{fig:7}(a)), this region extends from $\lambda\approx$ 0.4 to $\approx$0.75, while for $\Omega=2t$ (Fig.~\ref{fig:7}c) it extends from  $\lambda\approx$ 0.25 to $\approx$0.75. Therefore, the size of the metallic regime grows to extend over a larger range of $U$ and $\lambda$ as $\Omega$ increases. This trend was also found in studies of the one dimensional HH model.~\cite{clay} 

This $\Omega$-dependence arises because the $e$-$ph$ interaction more effectively renormalizes U$_{\mathrm{eff}}$ at larger phonon frequencies, as is evident from Eq.~\ref{eq:ueff}. Given that the size of the metallic regime depends on $\Omega$,  there are two ways that the system can transition between metallic and insulating states:  either by changing interaction strengths $\lambda$ and $U$ for fixed phonon frequency, or by varying $\Omega$ for fixed interaction strengths. The results in this section, along with the spin and charge susceptibilities in Sec.~\ref{susc}, support the insulator - metal - insulator evolution in Fig.~\ref{fig:1} by showing, using purely imaginary time quantities, the robustness of the insulating and metallic phases over a range of temperatures and phonon frequencies.

\section{Superconducting susceptibilities\label{sc}}
\begin{figure*}
\begin{center}
\includegraphics[width=0.9\textwidth]{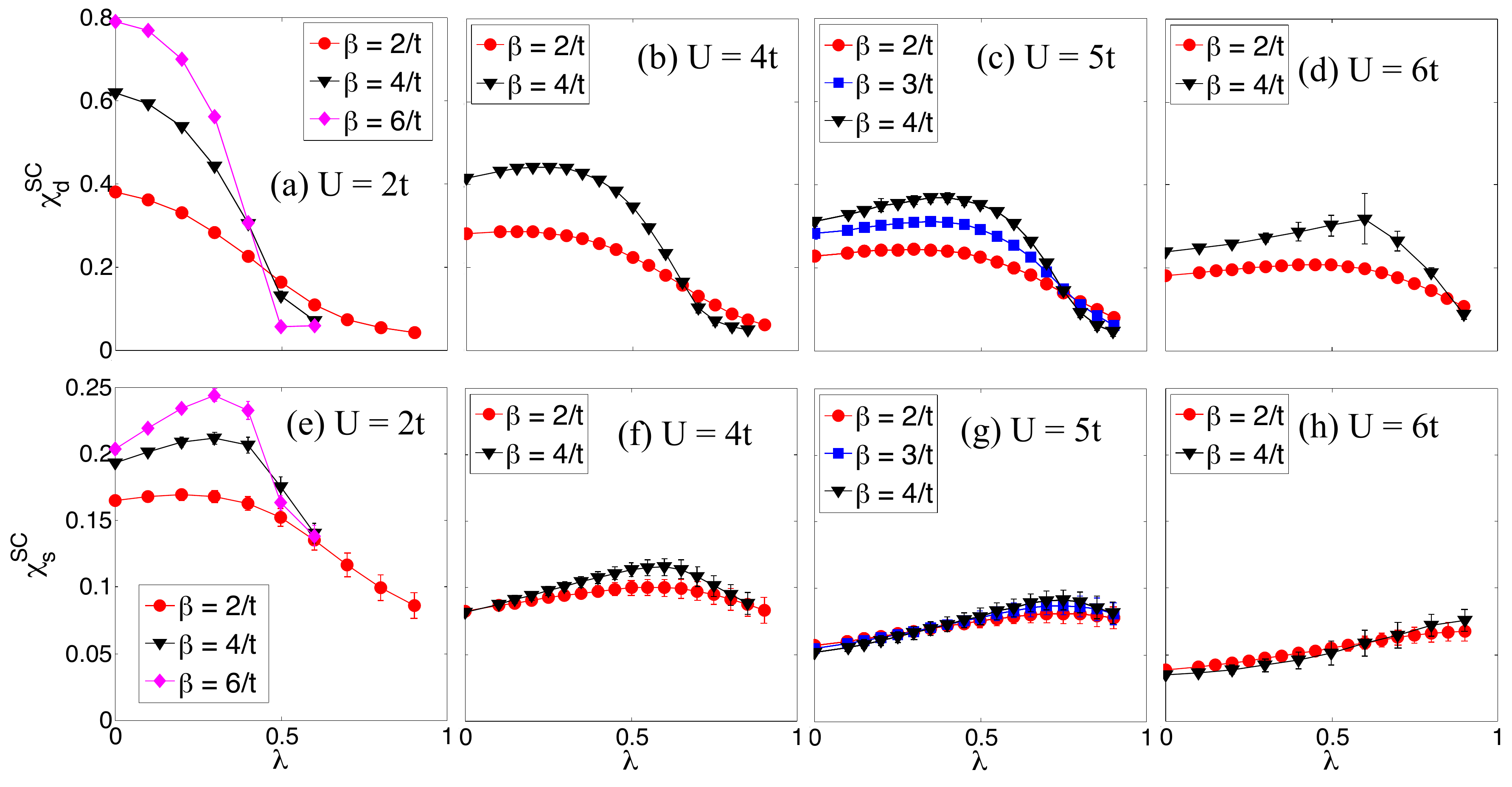}
\caption{\label{fig:12} $d$-wave pairing susceptibility at several temperatures for $U$= (a) $2t$, (b) $4t$, (c) $5t$, and (d) $6t$. $s$-wave pairing susceptibility for $U$= (e) $2t$, (f) $4t$, (g) $5t$, and (h) $6t$.  Other simulation parameters are $\Omega=t$ and $N=8\times 8$.}
\end{center}
\end{figure*}

Could the metallic phase become superconducting at lower temperature? Given the results in Figs.~\ref{fig:1} and ~\ref{fig:2}, the energy range for coherent quasiparticle formation introduces a new energy scale, in addition to the energy scales of the relevant superconducting pairing boson, be it phonon or spin fluctuations. With this  in mind, in Fig.~\ref{fig:12} we consider the superconducting pairing susceptibilities $\chi^{SC}_d$ and $\chi^{SC}_s$, which are defined as:
\begin{equation}
\chi^{SC}_{d,s} = \frac{1}{N}\int_0^\beta <T_\tau \Delta(\tau)\Delta^\dagger(0)>
\end{equation}
where for $d$-wave pairing 
\begin{equation}
\label{eq:dwave}
\Delta^\dagger = \frac{1}{2} \sum_{i\delta} P_\delta c_{i,\uparrow}^\dagger c^\dagger_{i+\delta,\downarrow}
\end{equation}
with the sum over $\delta$ running over nearest neighbor sites, and $P_{\pm \hat{x}} = 1 = -P_{\pm \hat{y}}$. For $s$-wave pairing,
\begin{equation}
\Delta^\dagger = \sum_i c_{i,\uparrow}^\dagger c_{i,\downarrow}^\dagger.
\end{equation}
Due to the fermion sign problem present in the HH model, we are unfortunately limited to relatively high temperatures so the superconducting susceptibilities are small in magnitude, however, their temperature dependence even in this regime may offer clues into their behavior at lower temperatures. 

We consider $\chi^{SC}_d$ and $\chi^{SC}_s$  in Fig.~\ref{fig:12}(a-d) and (e-h), respectively, for several values of $U$. First considering the $d$-wave pairing, $\chi^{SC}_d$ grows with lowering temperature for small to intermediate $\lambda$, while at strong $e$-$ph$ couplings, it is suppressed because the double-occupation of sites in the CDW phase is detrimental to the pair field defined in Eq.~\ref{eq:dwave}. 
 Interestingly, for $U$ = 5-6$t$, $\chi^{SC}_d$ peaks at an intermediate $\lambda$ value, suggesting that the $e$-$ph$ interaction enhances the $d$-wave pairing at moderate coupling strengths. The parameter ranges over which $\chi^{SC}_d$ grows with lowering temperature corresponds approximately to the metallic regime.
 
The $s$-wave pairing susceptibility is smaller in magnitude, and in particular, has a very weak temperature dependence, especially at larger $U$. It increases gradually as  $\lambda$ reduces the effective $U$, thus alleviating the suppression of  the on-site $s$-wave pairing. In summary, these results show a modest enhancement of the $d$-wave pairing in the metallic regime, which is consistent with the idea that the metallic phase becomes superconducting at low temperature, although access to lower temperatures would be necessary to make a more definitive statement.

\section{Conclusion \label{conclusion}}
In this paper, we analyzed the electron and phonon spectral functions of the half-filled HH model. With increasing $e$-$ph$ coupling, the spectral gap closes and later reopens as the system crosses between Mott insulating, metallic, and CDW insulating phases.  
The interplay of the $e$-$e$ and $e$-$ph$ interactions influences the low and high energy spectra in different ways. The high energy spectra become increasingly incoherent with increasing $\lambda$, while the effective interaction strength is suppressed at low energy, allowing  quasiparticles to  form. 

The phonon spectral function becomes momentum-dependent and the renormalized phonon frequency softens as the system approaches the CDW phase, although this is delayed to relatively strong $e$-$ph$ coupling by the presence of the $e$-$e$ interaction. We also study the temperature- and phonon frequency-dependence of the insulating and metallic phases, and find that the extent of the intervening metallic phase grows with increasing phonon frequency, in agreement with previous studies in one dimension. 

Finally, by considering superconducting pairing susceptibilities, we find a modest enhancement of the $d$-wave pairing strength in the parameter regimes corresponding to the metallic phase. Using alternative numerical methods that can access lower temperatures, an interesting direction for future studies would be to investigate whether the $d$-wave pairing becomes the dominant susceptibility in the metallic phase at low temperature. 

$Acknowedgements$: We acknowledge useful discussions with R. T. Scalettar and A. J. Millis. The computational work in this paper was partially performed at NERSC.  We acknowledge support from the U. S. Department of Energy, Office of Basic Energy Science, Division of Materials Science and 
Engineering under Contract No. DE-AC02-76SF00515. E. A. N. also acknowledges support from DOE Er-046169.

\appendix
\section{}
\begin{figure*}
 \begin{center}
\includegraphics[width=0.8\textwidth]{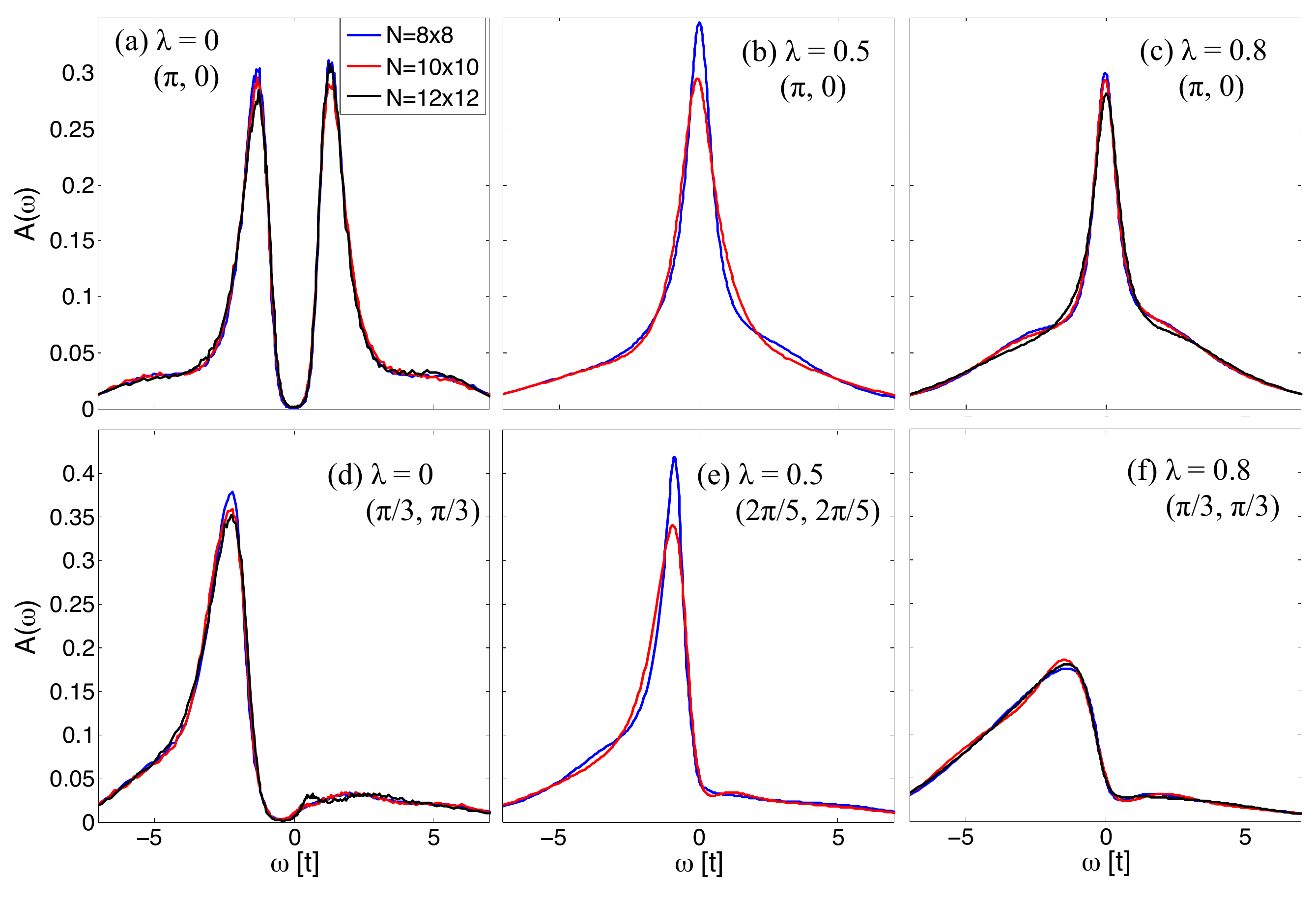}
\caption{\label{fig:10} Spectra obtained from clusters with different sizes $N$ at various momentum space points and $e$-$ph$ coupling strengths. Note that $(\pi, 0)$ is accessible from all three clusters considered, while the $N=8\times 8$ results in (d, e, f) and the $N=10\times 10$ results in (d, f) are interpolated.  The remaining simulation parameters are $U=6t$, $\Omega = t$, and $\beta=4/t$.}
\end{center}
\end{figure*}

In order to asses the influence of the cluster size on our spectra, in Fig.~\ref{fig:10} we compare spectra for $N = 8\times 8$, $N = 10 \times 10$, and $N = 12 \times 12 $ clusters. Fig.~\ref{fig:10} (a-c) show the  spectra at $(\pi, 0)$, which is a momentum space point accessible in all three clusters, for three different $e$-$ph$ coupling strengths $\lambda$. We find that the spectra from all three clusters are very similar, in particular, in the Mott insulating phase (Fig.~\ref{fig:10}a) the gap magnitude is identical for all clusters. Essentially the only difference between clusters is the quasiparticle peak height, which for the smaller clusters are systematically larger  than those from the $N=12\times 12$ case. Given the good agreement between spectra across these clusters, we conclude that the $N=8\times 8$ cluster used in the main part of this work contains sufficient momentum resolution to capture the important physics for the temperatures we consider.

To check the validity of the self-energy interpolation procedure described in Sec.~\ref{methods}, in Fig.~\ref{fig:10}d and f we consider the momentum space point $(\pi/3, \pi/3)$, which is directly accessible with the $N=12\times 12$ cluster, but requires interpolation from the momenta accessible in the $N=8\times 8$ and $N=10\times 10$ clusters. The interpolated spectra match very well with the $N=12\times 12$ result, with only minor differences in peak heights. In Fig.~\ref{fig:10}e, we consider $(2\pi/5, 2\pi/5)$ which is directly accessible with the $N=10\times 10$ cluster but requires interpolation for the $N=8\times 8$ case, and again find excellent agreement. These results confirm that interpolating the self-energy from an $N=8\times 8$ cluster to calculate spectra at arbitrary momenta yields accurate results.

We checked both the dependence of spectra on cluster size and interpolation for a variety of other momentum points, and found agreement consistent with the representative momentum space points shown here.

\bibliographystyle{apsrev}
\bibliography{nowadnick}

\begin{thebibliography}{57}
\bibitem[{foo({\natexlab{a}})}]{foot1}
\bibinfo{note}{If the phonon frequency were an order of magnitude smaller, so
  that the renormalizations from $e$-$e$ and $e$-$ph$ interactions occurred on
  different energy scales, we would expect to see two kink features in the
  spectra. However, given the momentum space resolution of cluster sizes
  accessible with DQMC, resolving a second kink at a substantially lower energy
  would be very difficult.}
\bibitem[{foo({\natexlab{b}})}]{foot2}
\bibinfo{note}{Due to noise in the data, numerical analytical continuation can
  produce slightly particle-hole asymmetric spectra. In Fig.~\ref{fig:2}(e-h)
  we plot $(\Sigma^\prime(\omega)-\Sigma^\prime(-\omega))/2$ to enforce
  particle-hole symmetry, and use these self-energies in the analysis in
  Fig.~\ref{fig:8}, although using the unsymmetrized self-energies produces a
  very similar result.}
\end{thebibliography}

\end{document}